\documentclass[a4paper, amsfonts, amssymb, amsmath, reprint, prapplied]{revtex4-2}
\usepackage[english]{babel}
\usepackage[utf8]{inputenc}
\usepackage[T1]{fontenc}
\usepackage{graphicx}
\usepackage{layouts}

\usepackage[pdftex, pdftitle={Article}, pdfauthor={Author}]{hyperref} 

\renewcommand{\rm}[1]{\mathrm{#1}}

\newcommand{\dd}[1]{\frac{\mathrm{d}}{\mathrm{d}{#1}}}

\renewcommand{\u}[1]{\,\mathrm{#1}}

\usepackage{xcolor}

\newcommand{\eref}{Eq.\,\eqref}

\newcommand{\fref}{Fig.\,\ref}

\newcommand{\tref}{Tab.\,\ref}

\begin{document}
\title{Optimizing the Cavity-Arm Ratio of V-Shaped Semiconductor Disk Lasers}

\author{Stefan Meinecke}
    \email[Correspondence email address: ]{meinecke@tu-berlin.de}
    \affiliation{Institut für Theoretische Physik, Technische Universität Berlin, Hardenbergstr. 36, 10623 Berlin, Germany}

\author{Kathy Lüdge}
    \email[Email address: ]{kathy.luedge@tu-ilmenau.de}
    \affiliation{Technische Universität Ilmenau, Institut für Physik, Weimarer Straße 25, 98693 Ilmenau, Germany}

\date{\today} 

\begin{abstract}

Passively mode-locked semiconductor disk lasers have received tremendous attention from both science and industry. Their relatively inexpensive production combined with excellent pulse performance and great emission wavelength flexibility make them suitable laser candidates for applications ranging from frequency comb tomography to spectroscopy. 
However, due to the interaction of the active medium dynamics and the device geometry, emission instabilities occur at high pump powers and thereby limit their performance potential. Hence, understanding those instabilities becomes critical for an optimal laser design.
Using a delay-differential equation model, we are able to detect, understand, and classify three distinct instabilities that limit the maximum achievable pump power for the fundamental mode-locking state and link them to characteristic positive net-gain windows. We furthermore derive a simple analytic approximation in order to quantitatively describe the stability boundary. Our results enable us to predict the optimal laser cavity configuration with respect to positive net-gain instabilities and are therefore of great relevance for the future development of passively mode-locking semiconductor disk lasers.

\end{abstract}


\maketitle

\section{Introduction}


Passively mode-locked semiconductor disk lasers generate regular sequences of short optical pulses without requiring any external modulation \cite{KEL03}. The research on and the engineering of such lasers has been subject to an outstanding progress for the past decades \cite{TIL15,WAL16,WAL18}. Nowadays, such lasers provide very competitive pulse performances, while offering flexible emission wavelengths and repetition rates due to their modular laser cavity design and the modern semiconductor band-gap engineering technologies \cite{TIL15}.
Pulse peak powers on the order of a few kilowatts \cite{WIL13b}, pulse widths as low as $100\u{fs}$ \cite{KLO11a}, and repetition rates ranging from $85\u{MHz}$ \cite{BUT13a} to $190\u{GHz}$ \cite{MAN14a,GAA16} have been reported.
Moreover, the mature semiconductor fabrication processes enable a relatively inexpensive production of semiconductor disk lasers, which makes them a popular choice for applications in frequency comb generation for metrology and spectroscopy \cite{UDE02,KEL06,KLE14,COD16,LIN17e,DID20}, super-continuum generation \cite{MAY15a,KLE16a,WAL16,WAL19}, and two-photon microscopy \cite{AVI11a,VOI17}. Since all those applications benefit from ultra short and high peak power pulses, the further optimization of passively mode-locked semiconductor disk lasers is highly desirable and an ongoing quest in both science and industry.

Typically, the achievable pulse peak power is limited by multi-pulse instabilities that arise at larger pump powers due to the relatively short upper state lifetimes of semiconductor gain media \cite{SAR14b}.
One approach to overcome this issue is to construct elaborate multi-pass geometries \cite{ZAU13,BUT13a,LIN17d,AVR19}, which prevent the built-up of an excess inversion in the gain chip and thus the formation of additional pulses.
The complex interplay between the gain medium dynamics and the laser cavity setup yields a fundamental mode-locking stability boundary, which characteristically depends on the geometric cavity configuration. Understanding this dependency therefore becomes crucial for an intelligent device design, which maximizes the laser performance.

\begin{figure}[htbp]
\centering
\includegraphics[width=\linewidth]{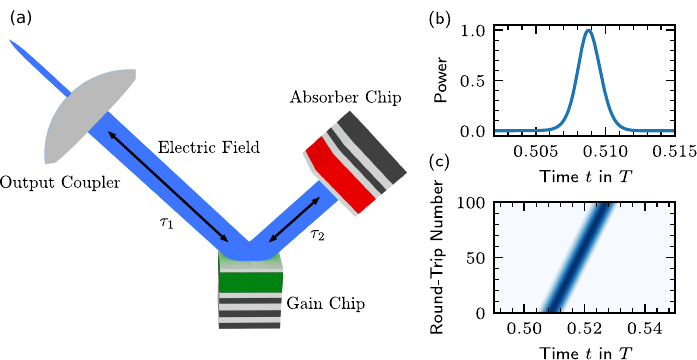}
\caption{(a) Sketch of the V-shaped external cavity laser. The setup contains three optical components: A high reflectivity output coupler, a semiconductor gain chip and a semiconductor saturable absorber chip. The two latter are mounted on top of high reflectivity DBRs. The active region of gain chip is indicated in green and the active region of the absorber in red.
The lengths of the cavity arms are denoted by their propagation times $\tau_1$ and $\tau_2$. (b) Pulse shape and (c) pseudo space-time plot of the fundamental mode-locking state.
}
\label{fig:setup}
\end{figure}

In this manuscript, we consider the optimization of a V-shaped external cavity, which represents one of the simplest multi-pass geometries. As sketched in \fref{fig:setup}\,(a), the output-coupler, gain chip, and absorber chip are arranged in V-shape, where the gain chip sits in the middle. Along one round trip through the cavity, the electric field therefore interacts twice with the gain chip and only once with the other two components. The passive free space in between the optical elements results in the field propagation time $\tau_1$ between the output coupler and the gain chip and the field propagation time $\tau_2$ between the gain chip and the absorber chip.
Such setups have been proven popular with semiconductor gain media and have thus been realized experimentally many times with examples given in Refs.\,\cite{HOO00,KLO11a,WIL13b,WAL16,GRO20,MEY20}. 
The particular device investigated in this manuscript is based on quantum-well active media and has been presented and characterized in detail in Refs.\,\cite{WAL16,WAL18}.

We investigate the impact of the geometric cavity configuration, i.e., the cavity-arm ratio, by the means of a minimalistic delay-differential equation model. The direct integration of the equations reveals the characteristic stability boundary of the, for applications relevant, fundamental mode-locking state. Moreover, the analysis of the temporal dynamics yields three distinct destabilizing mechanisms, which each can be fully understood in terms of the net-gain dynamics. We then use this insight to derive a simple analytic approximation for the net gain, which enables us to predict the stability boundary and thus to study the dependency on the other laser parameters. Our results allow us to predict the optimal cavity-arm ratio for passively mode-locked V-shaped lasers with respect to positive net-gain instabilities in terms of the maximum achievable pump power. Lastly, we discuss how our approach can also be applied to more complex gain media.

\section{Delay-Differential Equation Model}

To self-consistently describe the laser at the time scales required for a net-gain stability analysis, we use a multiple-delay differential equation model, which we have presented and derived in Refs.\,\cite{WAL18, HAU19, MEI22}. The model is obtained from a traveling-wave framework, which takes the forward and backward propagating light beams and the carrier dynamics inside the gain and absorber chips into account.
Integrating those quantities within the co-moving frame of the field propagation and applying a Lorentzian filter to model the gain spectrum yields delay-differential equations for the electric field $E$ at the output coupler and the integrated inversion densities $G$ and $Q$ for the gain and absorber chip \cite{HAU19,MEI22}:
\begin{alignat}{2}
 \dd{t} E(t) = & -\gamma E(t) + \gamma \sqrt{r_\rm{OC}} E(t-T) \nonumber \\
 & \times e^{\frac{1 - i \alpha_\rm{G}}{2}[G(t-\tau_1) + G(t-\tau_1-2\tau_2) ]} \nonumber\\
 & \times e^{(1-i \alpha_\rm{Q})Q(t-\tau_1 -\tau_2)} \nonumber\\
 & + \sqrt{D_\rm{sp}} \xi(t) \label{eq:vs_QW_DDE_E} \\
 \dd{t}G(t) = & - \gamma_\rm{G} G(t) + J_\rm{G} - \left( e^{G(t)} - 1 \right) \nonumber \\
 & \times \bigg[ \left|E(t-\tau_1)\right|^2 + \left|E(t-\tau_1-2\tau_2)\right|^2 \nonumber\\ 
 & \;\;\;\;\;\;\;\;\;\;\;\;\;\;\;\;\;\;\;\;\;\;\;\;\; \times e^{2Q(t-\tau_2)}e^{G(t-2\tau_2)}  \bigg] \label{eq:vs_QW_DDE_G} \\
 \dd{t}Q(t) = & \gamma_\rm{Q} (Q_0 - Q(t)) \nonumber\\
 &- s \left( e^{Q(t)} - 1 \right) \left|E(t-\tau_1-\tau_2) \right|^2 e^{G(t-\tau_2)}. \label{eq:vs_QW_DDE_Q}
\end{alignat}
The geometric cavity configuration, i.e., the length of the cavity arms, is then encoded via the propagation times $\tau_1$ and $\tau_2$ (s. \fref{fig:setup}), which appear in the time-delayed terms. Within the system equations, $\gamma$ describes the bandwidth of the gain medium, $r_\rm{OC}$ the output-coupler intensity reflectivity, $T$ the cold-cavity round-trip time, $\alpha_\rm{G}$ and $\alpha_\rm{Q}$ the amplitude-phase coupling coefficients of the gain and absorber chip, $\gamma_G$ the gain recovery rate, $J_\rm{G}$ the gain-chip pump current, $\gamma_Q$ the absorber recovery rate, $Q_0$ the unsaturated absorption and $s$ the ratio between the gain and absorber chip differential gain coefficients. Spontaneous noise is phenomenologically taken into account by a stochastic Langevin term in \eref{eq:vs_QW_DDE_E}, where $D_\rm{sp}$ describes the noise strength and $\xi(t)$ delta-correlation Gaussian white noise.

\begin{table}
\caption{\label{tab:parameters}Simulation parameters normalized to the cold-cavity round-trip time $T = 625\u{ps}$ unless stated otherwise.}
\begin{ruledtabular}
\begin{tabular}{lll}
Symbol & Value & Parameter  \\ \hline
$\gamma$ & 10000 & gain bandwidth \\
$r_\rm{OC}$ & 0.99 & output-coupler reflectivity \\
$\alpha_\rm{G,Q}$ & 0.0 & amplitude-phase coupling coefficient \\
$\tau_1$ & 0.25 & output-coupler - gain propagation time \\
$\tau_2$ & 0.25 & gain - absorber propagation time \\
$D_\rm{sp}$ & 0.1 & spontaneous emission noise strength \\
$J_\rm{G}$ & 0.04 & pump current \\
$\gamma_\rm{G}$ & 0.625 &  gain recovery rate \\
$Q_0$ & -0.03 & unsaturated absorption \\
$\gamma_\rm{Q}$ & 200.0 & absorber recovery rate \\
$s$ & 2.0 & differential gain ratio \\
\end{tabular}
\end{ruledtabular}
\end{table}

The model parameters are chosen to describe the laser investigated in \cite{WAL16,WAL18} and are given in \tref{tab:parameters}. However, we deliberately chose a narrower gain bandwidth $\gamma$ to stay within the slowly-varying envelope approximation and to enable fast simulation times. The resulting pulse widths are of the order of $\sim 1\u{ps}$ and thus still much shorter than the round-trip time $T$. Note that we set the amplitude-phase coupling coefficients to be zero, since the underlying approximations have been shown to break down at fast and high-power excitations \cite{AGR93,WAN07,HER16}, while the simulation of mode-locked lasers has been shown to be successful with vanishing amplitude-phase coupling \cite{PIM14,JAU16,NIK16}.
Nonetheless, we point out that our model at the chosen parameters does not include the phase dynamics of the electric field and subsequently can not describe any phase related instabilities. To investigate those, one should resort to more elaborate models \cite{KIL14, KIL16, PIM17, PIM19,HAU20a, MCL20}, which, however, come at the cost of increased computational demands.


To integrate the system equations, we utilize an explicit fourth-order Runge-Kutta algorithm \cite{PRE07} with cubic Hermite interpolation for the delay terms. The stochastic term in \eref{eq:vs_QW_DDE_E} is small at the chosen parameters and can thus be integrated alongside the deterministic equations via a simple Euler method.

\section{Results}

\subsection{Net-Gain Instabilities of the Fundamental Mode-Locking Regime}

\begin{figure}[htbp]
\centering
\includegraphics[width=\linewidth]{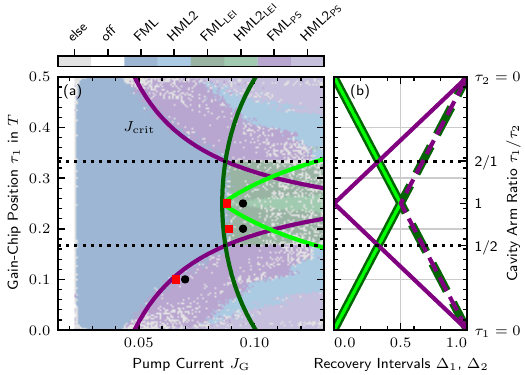}
\caption{Map of the emission states in the pump current and gain-chip position parameter space $(J_\rm{G}, \tau_1)$. (a) Numerically obtained emission state states with color-codes indicated on top. FML and HML denote fundamental and harmonic mode-locking and the subscripts LEI and PS denote leading-edge instability and pulse switching. Corresponding to the three instabilities, purple, dark green, and light green lines indicate the critical pump current $J_\rm{crit}$ obtained from the net-gain approximation \eref{eq:vs_Jcrit} with the effective pulse width $\Delta_\rm{pw} = 0.03$. Black circles indicate the parameter combinations used in \fref{fig:instab_space_time} and red squares indicate the parameter combinations used in \fref{fig:semistable}.
(b) Branches of the corresponding gain-recovery intervals $\Delta_1, \Delta_2$ (solid and dashed, respectively).
Horizontal black dotted lines indicate $\tau_1 = T/6$ and $\tau_1 = 2T/6$.
}
\label{fig:stability_bounds}
\end{figure}

The V-shaped semiconductors disk laser's underlying bifurcation structure of the relevant lasing solutions, albeit at broader pulses and larger gain parameters, has been published in Ref.\,\cite{HAU19}. The results reveal that the fundamental mode-locking state, which is the focus of this work, is born from a Andronov-Hopf bifurcation of the cw lasing state, similar to the cases of ring cavities \cite{VLA05, VLA10}. This bifurcation occurs very close to the lasing threshold, which can be determined analytically and reads
\begin{align}
	J_\mathrm{th} = \gamma_\rm{G} \left[\frac{1}{2}\ln{\left( \frac{\omega^2}{\gamma^2}+1\right)} -  \frac{1}{2}\ln{\left( \kappa \right)}-Q_0 \right], \label{eq:vs_Jth}
\end{align}
where $\omega$ denotes the maximum gain mode \cite{HAU19}. Note that the threshold does not depend on the cavity configuration $\tau_1, \tau_2$. The dynamics of the out-coupled field in the fundamental mode-locking state are exemplified in \fref{fig:setup}, where (b) shows the pulse shape and (c) a color-coded pseudo space-time plot. The smooth pulse shape remains stable from round trip to round trip, which demonstrates the stability of the chosen operation point.

To further investigate the impact of the cavity configuration on the fundamental mode-locking state, we numerically integrate the laser equations in the parameter space of the pump current and the cavity configuration. On that account, the overall cavity length, i.e., the cold-cavity round-trip time $T$, is kept constant while the ratio between the two cavity arms is changed. The geometric configuration is then specified in terms of the propagation time $\tau_1$, while the propagation time $\tau_2$ follows from the condition
\begin{align}
 2 \tau_1 + 2\tau_2 = T, \label{eq:vs_tau_T_cond}
\end{align}
where $T$ remains fixed as aforementioned. The propagation times thus take values $\tau_1, \tau_2 \in [0.0T,0.5T]$, where on the extremes, the gain chip is either located at the output coupler or at the absorber chip. These cases are not directly experimentally realizable with a V-shaped external cavity since the optical components take up some space by themselves. The intermediate configurations, however, can be easily achieved on an optical table by positioning the components accordingly.

The results are presented in \fref{fig:stability_bounds}\,(a), where the various lasing states are color-coded with the labels indicated on the top. While higher order states also exist, we want to focus our analysis in this manuscript onto the fundamental state and its instabilities.
Fundamental mode-locking (FML, blue) can be observed for all cavity configurations. As anticipated via the analytical threshold condition \eref{eq:vs_Jth}, the gain-chip position neither affects the lasing threshold nor the lower FML boundary. 
The upper pump current stability boundary, on the other hand, critically depends on the cavity configuration $\tau_1$. With the gain-chip located at the output coupler, i.e., $\tau_1 = 0.0$, the upper FML stability boundary is found at $J_\rm{G} \approx 0.048$ and pulse-switching unstable fundamental mode-locking (FML$_\rm{PS}$, dark purple) is found beyond the boundary. For an increasing gain-chip position $\tau_1$, the upper FML stability boundary shifts towards larger pump currents until it reaches a local maximum at $J_\rm{G} \approx 0.088$ for the configuration $\tau_1=1/6$.
From there on, the upper stability boundary marginally reduces to the local minimum $J_\rm{G} \approx 0.086$ at the symmetric configuration $\tau_1 = 0.25$. For cavity configurations $1/6 < \tau_1 < 0.25$, leading-edge unstable fundamental mode-locking (FML$_\rm{LEI}$, dark green) is observed beyond the stability boundary instead of FML$_\rm{PS}$. That is with the exception of the symmetric cavity configuration $\tau_1 = 0.25$, where leading-edge unstable second-order harmonic mode-locking (HML2$_\rm{LEI}$, light green) can be found beyond the stability boundary.
The remaining cavity configurations $\tau_1 \in \,]0.25,0.5]$ exhibit an apparent reflection symmetry with respect to $\tau_1 = 0.25$, which indicates that the output coupler and the absorber chip have roughly similar effects on the pulse shaping in fundamental mode-locking operation.

In conclusion, the cavity configuration has profound consequences on the possible fundamental mode-locking operation conditions. At the configurations $\tau_1 = 1/6$ and $\tau_1 = 2/6$, the range of available pump currents is $\approx 3.2$ larger than it is at $\tau_1 = 0.0$ and $\tau_1 = 0.5$, which moreover translates into maximally achievable pulse energies, which are $\approx 4.8$ times larger.

\begin{figure*}[htbp]
\centering
\includegraphics[width=\linewidth]{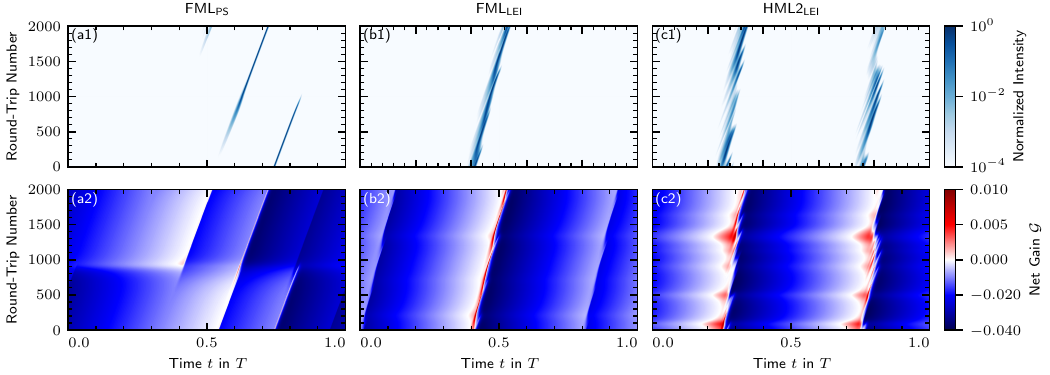}
\caption{Pseudo space-time plots of the normalized intensity $|E|^2$ (top row) and the net-gain $\mathcal{G}$ dynamics (bottom row).
From left to right: pulse-switching unstable fundamental mode-locking ((a), FML$_{\rm{PS}}$, $J_\rm{G} = 0.07$, $\tau_1 = 0.1$), leading-edge unstable fundamental mode-locking ((b), FML$_{\rm{LEI}}$, $J_\rm{G} = 0.095$, $\tau_1 = 0.2$), and leading-edge unstable second-order harmonic mode-locking ((c), HML2$_{\rm{LEI}}$, $J_G = 0.095$, $\tau_1 = 0.25$). The three parameter combinations are indicated by black circles in \fref{fig:stability_bounds}.
}
\label{fig:instab_space_time}
\end{figure*}

To unravel the mechanism, which leads to the characteristic fundamental mode-locking region, we take a closer look at the three different instabilities that can be observed beyond the stability boundary. Those are illustrated in \fref{fig:instab_space_time}, which shows color-coded pseudo space-time plots of the normalized intensity $|E|^2$ (top row) and the net gain $\mathcal{G}$ (bottom row). 
The columns from left to right present pulse-switching unstable fundamental mode-locking (FML$_{\rm{PS}}$), leading-edge unstable fundamental mode-locking (FML$_{\rm{LEI}}$), and leading-edge unstable second-order harmonic mode-locking (HML2$_{\rm{LEI}}$).

The net gain \cite{NEW74} measures the intensity amplification or attenuation, which a small perturbation experiences per round trip.
Evaluated for the V-shape laser model, the net gain contains the individual contributions of the various optical elements, which are picked up along the propagation and thus reads
\begin{align}
 \mathcal{G}(t) = & G(t-\tau_{1}) + G(t - \tau_{1} - 2\tau_{2}) \nonumber \\
                &+ 2Q(t - \tau_{1} - \tau_{2}) + \rm{ln}(\kappa). \label{eq:vs_netgain_2}
\end{align}
Positive net-gain, which does not coincide with the emission of a pulse, implies that perturbations can grow and thus may destabilize a regular mode-locking state. Note, however, that the net gain only represents a small signal measure that describes the short-term evolution of a perturbation that sits on top of existing lasing dynamics. Non-linear effects, e.g., mediated by the group-velocity dispersion, may allow mode-locking states to be stable despite positive net-gain windows \cite{VLA05,VLA11}. Nevertheless, the net gain will turn out to be the critical quantity to describe the observed instabilities of the fundamental mode-locking state.

Pulse-switching unstable fundamental mode-locking (FML$_\rm{PS}$) is characterized by pulse trains which only exist for a finite number of round trips before they vanish and a new pulse train is born at a different position within the cavity. As can be seen in \fref{fig:instab_space_time}\,(a2), the birth of a new pulse train is preceded by a broad positive net-gain window (white areas with a slight red tint). The new pulse then emerges at the trailing edge of the net-gain window.
As soon as the new pulse train has achieved sufficient power to effectively bleach the gain and absorber chip, it causes the positive net-gain window to shrink and only overlap with the pulse.
Shortly after the new pulse train is established, a new broad net-gain window establishes at different cavity position as well. Hence, the switching process regularly repeats and thereby prevents stable fundamental mode-locking emission. 

Leading-edge unstable fundamental mode-locking (FML$_\rm{LEI}$) is characterized by a single pulse train that exhibits a broad positive net-gain window at the leading edge of the pulse (s. \fref{fig:instab_space_time} (b2)). Hence, noise induced perturbations at the leading edge can be amplified and thus destabilize the pulse train. This produces pronounced fluctuations of the pulse peak intensities and of the pulse positions.

Lastly, leading-edge unstable second-order harmonic mode locking (HML2$_\rm{LEI}$) is characterized by two unstable and equidistant pulse trains. As shown in \fref{fig:instab_space_time}\,(c2), both pulses are accompanied by a large leading positive net-gain window, which continuously causes noise induced perturbations to grow and to destabilize the existing pulse trains. Being observed for a symmetric cavity configuration, the HML2$_\rm{LEI}$ state suffers from an additional destabilizing mechanism: The equidistant pulses collide in the gain chip, which tightly couples them via the interaction with the gain and thereby transfers instabilities from one pulse to the other.

In conclusion, the three different destabilizing mechanisms of the fundamental mode-locking state, which appear for different cavity configurations, are caused by characteristic positive net-gain windows. Those allow spontaneous emission noise to be amplified and consequently destabilize the regular mode-locking dynamics. We will see in the next section that this insight allows for an analytic treatment.

\subsection{Analytic Net-Gain Approximation}

To understand the shape of the stability boundary in the ($\tau_1,J_\rm{G}$) parameter plane (s. \fref{fig:stability_bounds}), we make further use of the net gain and derive an analytic approximation for the critical pump current of the three different instabilities.
The net gain $\mathcal{G}$ (s. \eref{eq:vs_netgain_2}) dynamically depends on the integrated gain $G$ and absorption $Q$. In the absence of optical pulses, both quantities, however, exhibit rather simple exponential relaxations. This occurs very quickly in the fast absorber chip, such that $Q(t)$ can be approximated by its unsaturated equilibrium value $Q_0$.
Assuming a complete saturation of the gain by the optical pulses, i.e., $G \approx 0.0$, the gain recovery can be analytically integrated and the result reads
\begin{align}
 G(\Delta) = \frac{J_\rm{G}}{\gamma_\rm{G}} \left( 1 - \exp(-\gamma_\rm{G} \Delta) \right),
\end{align}
where $\Delta$ denotes the time interval starting from the interaction of the pulse and the gain.
The net gain, in the absence of pulse interactions with the gain and absorber chip can then be written as
\begin{align}
 \mathcal{G}(\Delta_1, \Delta_2) = G(\Delta_1) + G(\Delta_2) + 2Q_0 + \rm{ln}(r_\rm{OC}), \label{eq:vs_netgain_approx}
\end{align}
where $\Delta_1$ and $\Delta_2$ denote the available gain-recovery time-intervals, which result from the last gain-chip interaction with the respective forward and backward travelling high-power pulse.
It is important to note that $\Delta_1$ and $\Delta_2$ depend on the emissions state and the cavity geometry as encoded by the gain-chip position $\tau_1$.
The previously introduced instabilities occur due to the appearance of distinct positive net-gain windows. This implies that our net-gain approximation in the absence of optical pulses \eref{eq:vs_netgain_approx} transitions to positive values for recovery intervals $\Delta_1, \Delta_2$, which are characteristic for the instability and the cavity configuration.
Hence, \eref{eq:vs_netgain_approx} is set to zero and solved for the critical pump current, which yields the expression
\begin{align}
 J_G^{\rm{crit}} = \frac{\gamma_\rm{G} \left(- \rm{ln}(r_\rm{OC}) - 2 Q_0 \right) }{ 2 - e^{-\gamma_\rm{G} [\Delta_1 - \Delta_\rm{pw}]} - e^{-\gamma_\rm{G} [\Delta_2 - \Delta_\rm{pw}] } }, \label{eq:vs_Jcrit}
\end{align}
where the recovery intervals $\Delta_1,\Delta_2$ are yet to be determined. Additionally, the recovery times have been adjusted for the finite pulse width $\Delta_\rm{pw}$, during which the electric field is not negligible and thus no free gain recovery occurs.

\begin{figure}[htbp]
\centering
\includegraphics[width=\linewidth]{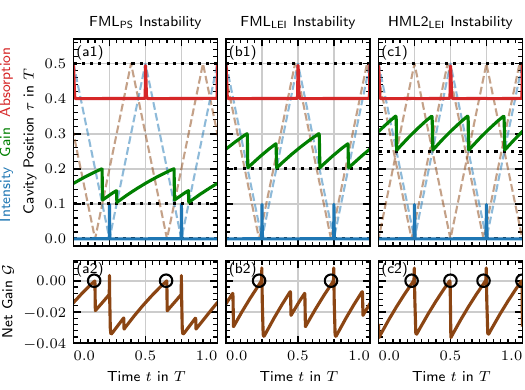}
\caption{Intensity and carrier dynamics of the fundamental mode-locking state close to the (a) pulse-switching instability ($J_{\rm{G}}=0.0675$, $\tau_1 = 0.1$), (b) leading-edge instability ($J_{\rm{G}}=0.0905$, $\tau_1 = 0.2$), and (c) second-order harmonic mode-locking leading-edge instability ($J_{\rm{G}}=0.0880$, $\tau_1 = 0.25$). Subfigures (a1) to (c1) show the temporal evolution of the intensity (blue lines), the gain (green lines), and the absorption (red lines), which have been normalized to $0.1$. Their respective zero positions (black dotted lines) have been shifted proportional to their cavity positions. The propagation of the pulses is illustrated by blue dashed lines and the propagation of the semistable perturbations by brown dashed lines.
(a2) to (c2) plot the respective net gain $\mathcal{G}$. Black circles indicate semistable ($\mathcal{G} \approx 0$) net gain windows. The three parameter combinations are indicated by red squares in \fref{fig:stability_bounds}.
 }
\label{fig:semistable}
\end{figure}

Each of the three instabilities produces an independent branch of recovery intervals $(\Delta_1, \Delta_2)$ that lead to positive net-gain windows. To identify the respective recovery intervals, the three different instabilities at their respective critical pump current $J_\rm{G}^\rm{crit}$ are further illustrated in \fref{fig:semistable}. 
The columns from left to right show the pulse-switching instability (a), the leading-edge instability (b), and the second-order harmonic mode-locking leading-edge instability (c).
The top and bottom row present the normalized intensity $|E|^2$ (blue), the gain $G$ (green), and the absorber $Q$ (red) dynamics ((a1) to (c1)) and the corresponding net-gain $\mathcal{G}$ dynamics ((a2) to (c2)). The zero baselines of the gain and absorption have been shifted proportional to their cavity position with respect to the output coupler and are indicated via horizontal black dotted lines. The field propagation through the laser cavity is illustrated via blue dashed lines. The intersections with the black dotted lines thus represent the interactions with the gain and absorber chips. Additionally, black circles in (a2) to (c2) indicate semistable ($\mathcal{G} \approx 0$) net-gain windows for which perturbations are neither amplified nor attenuated. The propagation of those perturbations is further indicated by brown dashed lines in (a1) to (c1). Note that the perturbations are considered to be small and thus leave no characteristic footprint on the gain dynamics, i.e., do not affect the gain recovery. On the other hand, the gain (green lines) is fully bleached by the optical pulses.

The recovery intervals $(\Delta_1, \Delta_2)$ can be identified as the time differences between a gain-chip interaction with the perturbation and the preceding high-power pulse. Crucially, the considered noise induced perturbations are small when they are generated, such that they can potentially be amplified twice by the same gain per round trip without causing any meaningful gain saturation.

The pulse-switching instability (FML$_\rm{PS}$) directly benefits from this mechanism. As \fref{fig:semistable}\,(a1) exemplifies, the backward-moving perturbation first passes the gain chip, is reflected at the output coupler, and then passes the gain chip in the forward direction right before the existing backward-moving pulse bleaches the gain. The associated recovery intervals read
\begin{align}
    \Delta_1 = 4 \left| \tau_1 - \frac{1}{4} \right|,\,\,\,\,\, \Delta_2 = \frac{1}{2} +  2 \left| \tau_1 - \frac{1}{4} \right|.
\end{align}
Note that the perturbation is amplified twice by the long gain-recovery, which leads to $\Delta_1, \Delta_2 \approx 1$ for configurations with $\tau_1 \approx 0.0$ and $\tau_1 \approx 0.5$.

The leading-edge instability (FML$_\rm{LEI}$), as illustrated in \fref{fig:semistable}\,(b1), is produced by a perturbation that travels right in front of the existing pulse. Hence, the characteristic recovery intervals correspond to the propagation times that determine the pulse amplification in \eref{eq:vs_QW_DDE_E} and read
\begin{align}
    \Delta_{1,2} = \frac{1}{2} \pm 2 \left| \tau_1 - \frac{1}{4} \right|
\end{align}

Lastly, the second-order leading-edge instability (HML2$_\rm{LEI}$) is generated by a perturbation that reaches the gain chip in the backward direction right before the existing forward-moving pulse saturates the gain, as shown in \fref{fig:semistable}\,(c1). This produces identical recovery intervals
\begin{align}
    \Delta_{1,2} &= \frac{1}{2} - 2 \left| \tau_1 - \frac{1}{4} \right|.
\end{align}
Note that both recovery intervals drop to zero for the configurations $\tau_1 = 0.0$ and $\tau_1 = 0.5$, which causes \eref{eq:vs_Jcrit} to diverge.

The three branches of gain-recovery intervals $\Delta_{1,2}$ are plotted as functions of the cavity configuration $\tau_1$ in \fref{fig:stability_bounds}\,(b), where solid lines denote $\Delta_{1}$ and dashed lines denote $\Delta_{2}$.
Purple represents the pulse-switching instability (FML$_\rm{PS}$), dark green the leading-edge instability (FML$_\rm{LEI}$), and light green the second-order leading-edge instability (HML2$_\rm{LEI}$). The corresponding critical pump currents $J_\rm{G}^\rm{crit}$ are computed via \eref{eq:vs_Jcrit} and plotted in \fref{fig:stability_bounds}\,(a).

As demonstrated by the comparison to the underlying numerically obtained emission states, the analytic approximation \eref{eq:vs_Jcrit} excellently predicts the fundamental mode-locking upper stability boundary as well as the respective destabilizing mechanism that appears beyond the critical pump current. On that account, fundamental mode-locking stability is determined by the lowest critical pump current $J_\rm{G}^\rm{crit}$ among the three instabilities. 
For the rather asymmetric configurations, i.e., $\tau_1 < 1/6$ and $\tau_1 > 2/6$, this turns out to be the pulse-switching instability (purple line). According to \fref{fig:stability_bounds}\,(b), this can be attributed to the longest available recovery intervals $\Delta_1$ and $\Delta_2$, which already enable smaller pump currents to achieve positive net-gain windows.

For rather symmetric configurations $1/6 < \tau_1 < 2/6$, the leading-edge instability (dark green lines in \fref{fig:stability_bounds}) provides the longest recovery intervals $(\Delta_1, \Delta_2)$ and thus first destabilizes the fundamental mode-locking state.
The pump current stability boundary maxima at $\tau_1 = 1/6$ and $\tau_1 = 2/6$ are generated by the intersection of the pulse-switching (purple line) and leading-edge instability (dark green line). At those points, the device geometry yields identical recovery intervals $\Delta_{1,2}$ for both instabilities (s. \fref{fig:stability_bounds}\,(b)).
Moreover, the leading-edge instability exhibits a global minimum for the symmetric cavity $\tau_1 = 0.25$, despite the constant sum $\Delta_1 + \Delta_2 = 1$ of the two recovery intervals. This feature is caused by the concave downward recovery of the gain $G$, which is more efficient, i.e., faster, for shorter recovery intervals and thus requires a smaller critical pump current to produce positive net-gain windows.

Lastly, the second-order leading-edge instability (light green lines) becomes only relevant for the symmetric cavity configuration $\tau_1 = 0.25$, where the recovery intervals  $(\Delta_1, \Delta_2)$ are identical to the leading-edge instability. For all other configurations, the critical pump current quickly grows and diverges towards $\tau_1 = 0.0$ and $\tau_1 = 0.5$. Hence, the leading-edge unstable second-order harmonic mode-locking state HML2$_\rm{LEI}$ can only be found right beyond the stability boundary for the symmetric configuration as well as in a small cone, which grows in the $(J_\rm{G}, \tau_1)$ parameter space for further increasing pump currents $J_\rm{G}$.

\begin{figure}[htbp]
\centering
\includegraphics[width=\linewidth]{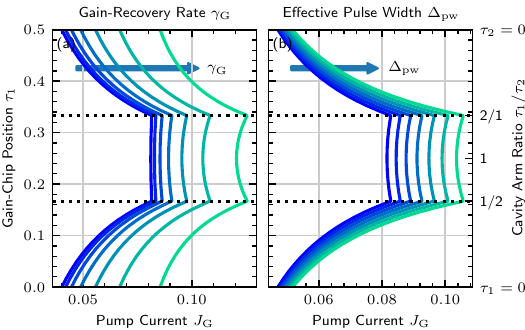}
\caption{Critical pump current $J_{\rm{G}}^{\rm{crit}}$ (upper stability boundary) of the fundamental mode-locking state in the pump current and gain-chip position parameter space $(J_\rm{G}, \tau_1)$. Subfigure (a) plots $J_{\rm{G}}^{\rm{crit}}$ for a geometrically increasing gain-chip recovery rate $\gamma_\rm{G} \in \{0.185,\ldots,2.11\}$. Subfigure (b) plots $J_{\rm{G}}^{\rm{crit}}$ for a linearly increasing pulse width $\Delta_\rm{pw} \in \{0.0,\ldots,0.12\}$.
}
\label{fig:netgain_approx}
\end{figure}

In summary, the presented results demonstrate that the stability boundaries of the fundamental mode-locking state can be entirely understood in terms of the net-gain dynamics and can be analytically predicted via a simple gain-recovery approximation. Lastly, we now utilize this approximation to study the relevant parameter dependencies of the upper FML stability boundary. The critical pump current $J_G^{\rm{crit}}$ given in \eref{eq:vs_Jcrit} linearly scales with the logarithm of the out-coupling  losses  $r_\rm{OC}$ and the unsaturated absorption $Q_0$ and nonlinearly with the gain recovery rate $\gamma_\rm{G}$ and the effective pulse width $\Delta_\rm{pw}$.
The influence of the two latter in the pump current and gain-chip position parameter space $(J_\rm{G}, \tau_1)$ is presented in \fref{fig:netgain_approx}. The critical pump current $J_G^\rm{crit}$ is determined as the minimum among the three destabilizing mechanisms and seven representative stability boundaries are plotted, respectively.

The gain-recovery rate $\gamma_\rm{G}$, shown in \fref{fig:netgain_approx}\,(a), is increased geometrically (constant multiplication factor) from $\gamma_\rm{G} = 0.185$ to $\gamma_\rm{G} = 2.11$. The increasing recovery rate most dominantly shifts the critical pump current to larger values due to the linear factor of $\gamma_\rm{G}$ in the numerator of \eref{eq:vs_Jcrit}. Secondly, a fast (large) gain-recovery rate produces an effectively 'more concave' gain recovery via the contributions in the exponentials in the denominator of \eref{eq:vs_Jcrit}. This leads to more pronounced local maxima and a stronger curvature of the stability boundary at fast recovery rates. At small rates, the recovery is slow and effectively linear at the available recovery intervals $\Delta_1,\Delta_2$, which leads to a leading-edge instability boundary that is almost constant between gain-chip positions $1/6 \leq \tau_1 \leq 2/6$.

The effective pulse width $\Delta_\rm{pw}$, shown in \fref{fig:netgain_approx}\,(b), is increased linearly with equidistant steps from $\Delta_\rm{pw} = 0.0$ to $\Delta_\rm{pw} = 0.12$. The effective pulse width only appears in the exponentials in \eref{eq:vs_Jcrit} and reduces the effective gain recovery intervals $\Delta_1$ and $\Delta_2$. Hence, it increases the pump current $J_\rm{G}$, which is needed to achieve positive net gain. The effect is nonlinear and, similarly to the gain-recovery rate, leads to a more pronounced curvature and stronger local maxima at larger pulse widths (light green curve to the right in (b)). 

In conclusion, the nonlinear dependencies of the critical pump current determine the curvature of the stability boundary. Nonetheless, the local maxima are always located at $\tau_1 \approx 1/6$ and $\tau_1 \approx 2/6$. However, their relative advantages in terms of the pump current tuning range with respect to neighboring cavity configurations increases for a stronger nonlinearity of the gain-recovery process. Note that the relative nonlinearity of the gain recovery can also be controlled via the scaling of the cavity size, since the presented discussion was performed with parameters normalized to the cold-cavity round-trip time.

\section{Discussion}

We have demonstrated that the stability of the fundamental mode-locking state of a V-shaped semiconductor disk laser can be qualitatively and quantitatively understood in terms of its net-gain dynamics. In particular, the maximum achievable pump current is limited by positive net-gain instabilities, which, depending on the geometric cavity configuration, manifests either as a pulse switching, leading edge, or second-order leading-edge instability.
The corresponding stability boundaries of those three destabilizing mechanisms can be analytically predicted by solving the gain recovery in the absence of optical pulses and using the results to construct the net gain as a function of the available gain-recovery intervals.

We have associated the three instabilities each with a branch of gain-recovery times $(\Delta_1,\Delta_2)$. Those depend on the cavity-arm ratio and thereby determine, which instability occurs at the lowest pump current and thus limits the maximum pump current. The branches $(\Delta_1,\Delta_2)$ intersect at characteristic cavity configurations, which in turn causes the corresponding stability boundaries to also intersect. This produces the characteristic region of stable FML emission in the pump current and cavity configuration parameter space, which exhibits the optimal pump current tuning range for the cavity-arm ratios $\tau_1/\tau_2 = 1/2$ and $\tau_1/\tau_2 = 2/1$.
We have furthermore observed that the region of stable FML emission retains its qualitative features, i.e., its shape, for all changes of the relevant laser parameters. We attribute those results to the concave downward gain-recovery process and the cavity configuration, which determines the available gain-recovery intervals.

For that reason, the qualitative shape of the stable fundamental mode-locking region should generalize to any gain-chip active medium, which exhibits a concave downward recovery process. Hence, we predict that the largest pump current tuning range should always be observed at the characteristic cavity-arm ratios $\tau_1/\tau_2 = 1/2$ and $\tau_1/\tau_2 = 2/1$. 
However, for more complex active media, the gain dynamics can likely not be solved analytically, which would prevent a simple approximation of the stability boundaries. Nevertheless, to obtain quantitative predictions, the stability boundary condition ($\mathcal{G} \approx 0$) could be formulated as an ordinary-differential equation boundary problem. The numerical solution of this problem should be always much cheaper than the simulation of the complete spatio-temporal laser dynamics. 
This approach could be easily implemented to calculate the stability boundaries for gain-chips with quantum-dot \cite{LUE11b,HOF11a,LIN14,MEI19}, submonolayer quantum-dot \cite{HER16,ALF18,HAU21}, and multi-level quantum-well active media \cite{ALF17,WAL18}.

\section{Achknowledgements}
We thank Jan Hausen for fruitful discussions and pioneering the bifurcation analysis of the V-shaped laser model.

\section{Code Availability}
The simulation code is available on GitHub under a MIT license (\url{https://github.com/stmeinecke/VShape})

\end{document}